# Two distinct approaches to tune multiple reflective bands of one-dimensional photonic crystal at normal incidence

Quan-Shan Liu[1,2], Lu Qiu[1,2], Tao Wen[1,2]*, Rui Zhang[1,2]*

[1]South China Advanced Institute for Soft Matter Science and Technology, School of Molecular Science and Engineering, South China University of Technology, Guangzhou 510640, China.
[2]Guangdong Provincial Key Laboratory of Functional and Intelligent Hybrid Materials and Devices, South China University of Technology, Guangzhou 510640, China.

*Corresponding authors: rzhang1216@scut.edu.cn (R.Z.); twen@scut.edu.cn (T.W.).

**ABSTRACT**:
One-dimensional photonic crystals (1D PC) represent a class of periodic optical material, composed of alternating media with different dielectric constants along one direction. The most important property of 1D PCs is their photonic band-gap. However, multiple reflective bands are rarely reported in this research area. In this paper we demonstrate the tunability of multiple reflective bands in conventional 1D PC structure and 1D PC heterostructure. For both two types of 1D PC construction, positions of multiple reflective bands can be regulated under certain principles. In addition, structural color is revealed by transforming reflection spectra into CIE coordinates. It is indicated that the CIE coordinate shifts caused by multiple reflective bands behave quite different compared to those caused by one major photonic band-gap. The two approaches reported in this work may provide insights for the application of 1D PC in areas such as displays, sensors, and decoration.

## 1. Introduction

Photonic crystals (PCs) have been an eye-catching research topic in photonics science since the concept was first proposed [1,2]. The most striking feature of PCs is the photonic bandgap, where light falling in a certain range of wavelengths cannot propagate freely. Among the family of PCs, one-dimensional (1D) PCs are composed of alternating dielectric layers along one direction, which characterizes them as simple structures and convenient fabrication [3,4]. Besides, taking advantage of vivid structural color originating from periodic arrangements [5–9], 1D PCs are increasingly applied in areas such as sensing technology [10–13], anti-counterfeit materials [14,15], and light-emitting devices [16–18]. Usually, the variability of structural color can be realized by tuning the first-order reflective band of 1D PCs. Intuitively, manipulation of structural color by using multi-order photonic reflections is intriguing and may give rise to unprecedented colors [19]. However, effective utilization of multiple reflective bands in 1D PC systems is rarely reported [20–22].

Herein, we report two distinct approaches to tune multiple reflective bands of 1D PCs in the visible light region. The two approaches are applicable to two different kinds of 1D PC construction respectively. Except for conventional 1D PC architecture, another type of 1D PC construction considered in the present work is 1D PC heterostructure [23,24]. To the best of our knowledge, the rational design of multiple reflective bands towards structural color has not been thoroughly discussed in previous literature, and we hope this report may provide a new vision for 1D PC research.

## 2. Model and Method

2.1 Conventional 1D PC architecture

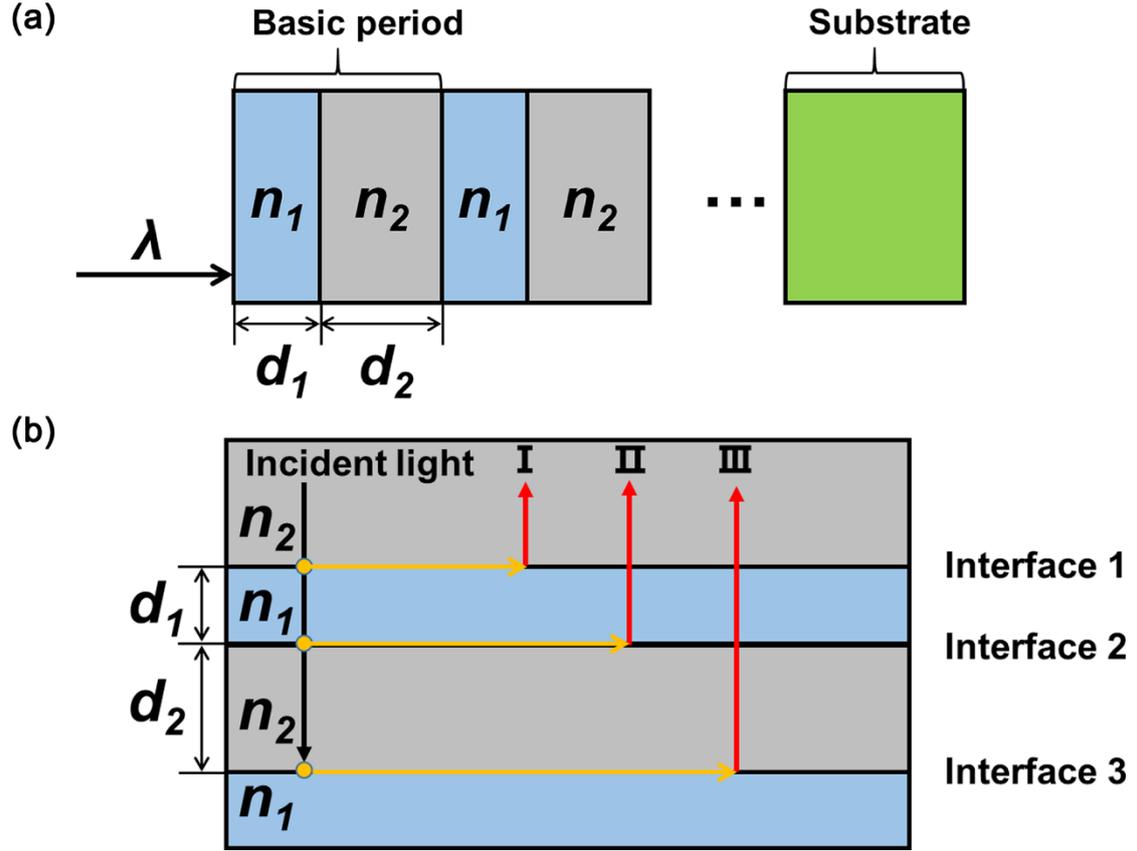

FIG. 1. (a) Schematic configuration of a conventional 1D PC with a substrate. (b) Schematic of light path occurring in one basic period. Three beams of reflected light are marked as I, II, and III respectively. Orange lines are used to differentiate the reflected beams, and they do not represent real light paths.

Fig. 1(a) shows the configuration of conventional 1D PC, which is composed of alternating dielectric materials with different refractive indices $n_1$ and $n_2$. Here, $d_1$ and $d_2$ represent the thicknesses of individual layers respectively. In literature, the quarter-wave condition is usually adopted for theoretical guidance [25]. According to this condition, the first-order reflective band at normal incidence is maximized when $n_1d_1 = n_2d_2$, and it will appear at $\lambda_{1st} = 2(n_1d_1+n_2d_2)$. The case is more interesting when $n_1d_1 \neq n_2d_2$ in that we can modulate the distribution of normal-incidence multiple reflective bands [26].

To illustrate how multiple reflective bands can be tuned, we start the analysis from one basic period, as shown in Fig. 1(b). Firstly, it is assumed that $n_1 > n_2$ and multiple reflected beams between adjacent interfaces are ignored. It can be deduced that when beam I and beam III are in phase, there will be a large number of beams contributing to constructive interference as the basic period repeats, resulting in multiple reflective bands. Half-wave loss occurs both on interface 1 and interface 3, so the effect is offset when calculating the optical path difference between beam I and beam III. Thus, the necessary condition of forming multiple reflective bands can be given in the following equation:

$$2(n_1 d_1 + n_2 d_2) = a \times \lambda \tag{1}$$

where $a$ is some positive integer and $\lambda$ is some certain wavelength. A series of central wavelengths can be obtained as $a$ changes. However, equation (1) is not a sufficient condition for forming multiple reflective bands. Provided beam I and beam III are in-phase, we proceed to consider the phase relation between beam I and beam II. Here, we need to avoid $2n_2 d_2 = b \times \lambda'$, where $b$ is some positive integer and $\lambda'$ is some certain wavelength. The reason is that the effect of additional half-wave loss converts constructive interference into destructive interference, leading to the weakness of some reflective bands. Thus, the sufficient and necessary condition of forming multiple reflective bands can be expressed as follows:

$$2(n_1 d_1 + n_2 d_2) = a \times \lambda \ \& \ 2n_2 d_2 \neq b \times \lambda' \tag{2}$$

In the case of $n_1 < n_2$, the same result can be obtained by a similar deduction. Based on equation (2), we discuss three scenarios: (i) $n_1 d_1 = n_2 d_2$. Then equation (2) becomes $4n_2 d_2 = a \times \lambda \ \& \ 2n_2 d_2 \neq b \times \lambda'$. Therefore, multiple reflective bands will be located at $\lambda_{1st}$, $\lambda_{1st}/3$, $\lambda_{1st}/5$ ⋯. To conclude, the corresponding divisors cannot be multiples of 2. (ii) $n_1 d_1 = 2n_2 d_2$. Then equation (2) becomes $6n_2 d_2 = a \times \lambda \ \& \ 2n_2 d_2 \neq b \times \lambda'$. As a result, multiple reflective bands will be located at $\lambda_{1st}$, $\lambda_{1st}/2$, $\lambda_{1st}/4$, $\lambda_{1st}/5$, $\lambda_{1st}/7$ ⋯. To sum up, the corresponding divisors cannot be multiples of 3. (iii) $n_1 d_1 = 3n_2 d_2$. Then equation (2) becomes $8n_2 d_2 = a \times \lambda \ \& \ 2n_2 d_2 \neq b \times \lambda'$. Consequently, multiple reflective bands will be located at $\lambda_{1st}$, $\lambda_{1st}/2$, $\lambda_{1st}/3$, $\lambda_{1st}/5$, $\lambda_{1st}/6$, $\lambda_{1st}/7$, $\lambda_{1st}/9$ ⋯. To summarize, the corresponding divisors cannot be multiples of 4.

Practical examples will be given in chapter 3.

2.2 1D PC heterostructure

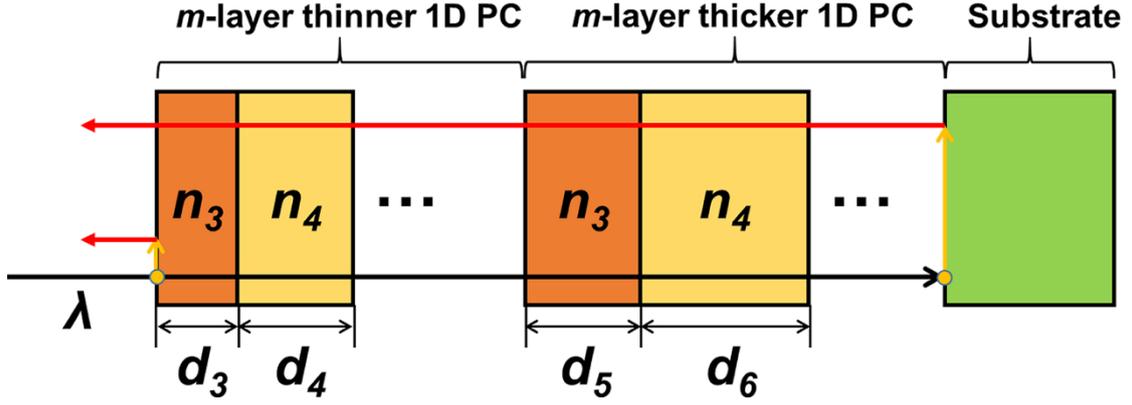

FIG. 2. Schematic configuration of a 2*m*-layer 1D PC heterostructure, which is constructed by two different 1D PCs and a substrate. Two specified beams of reflected light are denoted by red lines, while orange lines are used to differentiate the reflected beams, and they do not represent real light paths.

Fig. 2 shows the configuration of the 1D PC heterostructure. Here, *n* and *d* refer to refractive index and layer thickness respectively, and 2*m* is the number of dielectric layers of the heterostructure. Besides, *m* has to be some positive even number so as to retain the periodicity of individual 1D PC. As shown in Figure 2, the two different 1D PCs are comprised of the same dielectric materials, but with distinct layer thicknesses. There are two forms of heterostructure depending on the stacking order of the two 1D PCs. In this paper, we adopt the model of air/*m*-layer thinner 1D PC/*m*-layer thicker 1D PC/substrate. It was reported that by maintaining low refractive index contrast between dielectric layers and taking a substrate with a high refractive index, multiple reflective bands can be tuned by controlling the optical thickness of the heterostructure [22]. Positions of multiple reflective bands can be predicted by the following equation:

$$c \times \lambda'' = 2 \sum_i n_i d_i \qquad (3)$$

where *c* is some positive integer and *λ"* is the central wavelength of the corresponding reflective band. The summation is performed for all dielectric layers except for the substrate, and prefactor 2 links the optical thickness to the optical path difference (OPD) between two specified beams of reflected light, as illustrated in Fig. 2. However, the optical thickness can only give information about band positions, and the reflectance of the individual reflective band is still unknown.

There are various material thickness combinations for the two constituent 1D PCs when keeping the optical thickness of the heterostructure fixed. For the two 1D PCs, the positions of the respective first-order reflective bands can be calculated using equation (1) by setting $a = 1$. It was indicated that reflective bands near the two positions have higher reflectance [22]. Based on the above indication, two suggestions can be made: one is that the two positions are supposed to be tuned to approach reflective bands of interest, and the other is that the optimum number of dielectric layers can be chosen according to the orders of reflective bands of concern. The second suggestion is actually an auxiliary condition contributing to the realization of the first suggestion.

To make it clearer, we start from one example where fifth, sixth, and seventh-order reflective bands are studied. Firstly, the order $c$ is set to be 6, and an initial attempt of the number of dielectric layers $2m$ is 12. By simplifying equation (3), the final result can be expressed as $\lambda''_{6th} = 2(n_3d_3+n_4d_4+n_3d_5+n_4d_6)/2$. It can be seen that the sixth reflective band is located in the middle of the two positions. In this case, the two positions can be well adjusted to approach the three reflective bands. Next, we consider the cases of $2m = 8$ and $2m = 16$. Then equation (3) can be simplified as $\lambda''_{6th} = 2(n_3d_3+n_4d_4+n_3d_5+n_4d_6)/3$ and $4(n_3d_3+n_4d_4+n_3d_5+n_4d_6)/3$ respectively. In these two cases, it would be difficult to keep both two positions adjacent to the three reflective bands. It can also be inferred that the situation won't be better when $2m$ takes other values. The whole analysis process is straightforward and can be applied to other circumstances.

Concrete examples will be introduced in chapter 3.

2.3 Simulation method and materials

All the simulation in this paper was conducted at normal incidence using the transfer matrix method [27]. During the simulation, we consider the following dielectric materials $TiO_2$, $SiO_2$, titania sol, and UV-curable resin, with their refractive indices adopted to be 2.34, 1.46 [20], 1.78, and 1.54 [28] respectively. They are all transparent in the visible light region, so light absorption loss is ignored. Glass and silicon wafer (Si) are chosen as substrate materials, with refractive indices as 1.52 and 3.90 respectively.

## 3. Results and discussion

3.1 Yellow and blue double-mode reflective bands in the visible region

In this section, conventional 1D PC architecture and 1D PC heterostructure are used to provide yellow and blue double-mode reflective bands in the visible region. For the conventional 1D PC structure, we chose TiO$_2$ ($n_1$ = 2.34) and SiO$_2$ ($n_2$ = 1.46) as dielectric materials, and the number of dielectric layers was set to be 8. Based on scenario (ii) in section 2.1, we focused on fourth- and fifth-order reflective bands. Regarding the relationship between material thicknesses, we adopted that $n_1d_1 = 2n_2d_2 = (4/3) \times \lambda_{1st}/4$, where $\lambda_{1st}$ equals 2300nm. The theoretical reflection spectra for conventional 1D PC deposited on different substrates were obtained, labeled as "D-glass" and "D-Si" respectively, as shown in Fig. 3(a). Here, "D" is short for "double-mode". It can be seen that two high reflective bands are well situated in yellow and blue wavebands respectively. Another fact is that the change of substrate has a limited impact on the band shapes of the two high reflective bands.

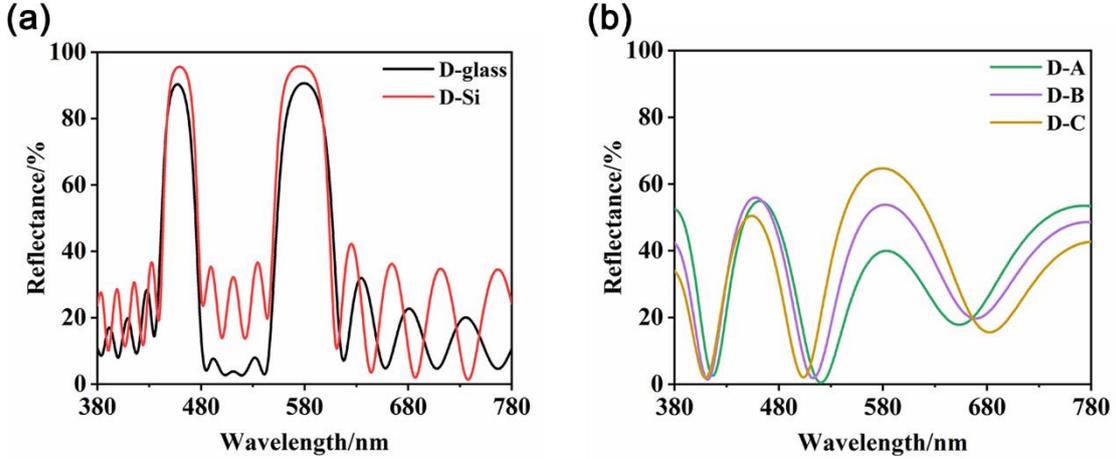

FIG. 3. (a) Reflectance spectra of conventional 1D PC on different substrates. (b) Reflectance spectra of 1D heterostructures with different material thickness combinations.

For the 1D PC heterostructure, we chose titania sol ($n_3$ = 1.78) and UV-curable resin ($n_4$ = 1.54) as dielectric materials, with Si served as the substrate. Based on the second suggestion in section 2.2, the number of dielectric layers was set to be 8. The OPD described in equation (3) was restricted to be 2300nm. Regarding layer thicknesses, the quarter-wave condition was adopted for each constituent 1D PC. Under this premise, we only need to assign two parameters, which are the positions of respective first-order reflective bands of the two 1D PCs. For convenience, the two positions are symbolized by $\lambda_1$ and $\lambda_2$ respectively, and exact layer thicknesses are expressed in the formulation of $\{\lambda_1/4n_3, \lambda_1/4n_4, \lambda_2/4n_3, \lambda_2/4n_4\}$. While the OPD is constant, material thickness combinations can be diverse. Here, three groups of material thicknesses were considered, denoted by "D-A" with $\{420/4n_3, 420/4n_4, 730/4n_3, 730/4n_4\}$, "D-B" with $\{460/4n_3, 460/4n_4, 690/4n_3, 690/4n_4\}$, and "D-C" with $\{500/4n_3, 500/4n_4, 650/4n_3, 650/4n_4\}$, respectively. Fig. 3(b) exhibits the reflection spectra of the three heterostructures. It is suggested that the relative intensities of the two reflective bands can be controlled by changing material thickness combinations.

3.2 Red, green, and blue triple-mode reflective bands in the visible region

In this section, conventional 1D PC architecture and 1D PC heterostructure are used to provide red, green, and blue triple-mode reflective bands in the visible region. For the conventional 1D PC structure, we chose $TiO_2$ ($n_1 = 2.34$) and $SiO_2$ ($n_2 = 1.46$) as dielectric materials, and the number of dielectric layers was set to be 12. Based on scenario (iii) in section 2.1, we focused on fifth, sixth, and seventh-order reflective bands. Regarding the relationship between material thicknesses, we adopted that $n_1d_1 = 3n_2d_2 = (3/2) \times \lambda_{1st}/4$, where $\lambda_{1st}$ equals 3150nm. The theoretical reflection spectra for conventional 1D PC deposited on different substrates were obtained, labeled as "T-glass" and "T-Si" respectively, as illustrated in Fig. 4(a). Here, "T" is short for "triple-mode". Fig. 4(a) reveals that three high reflective bands are accurately located in red, green, and blue wavebands respectively. It can also be observed that the effect of the change of substrate on the band shapes of the three high reflective bands is almost negligible.

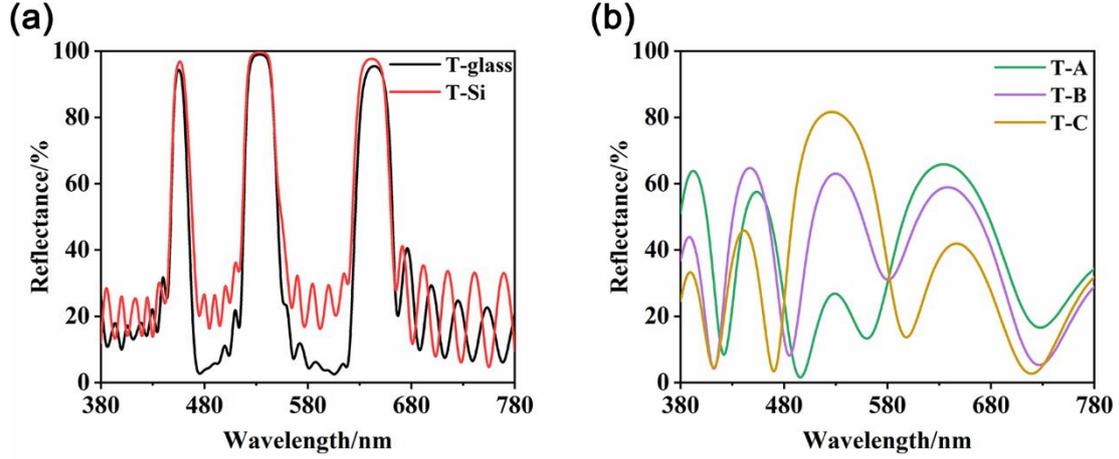

FIG. 4. (a) Reflectance spectra of conventional 1D PC on different substrates. (b) Reflectance spectra of 1D heterostructures with different material thickness combinations.

For the 1D PC heterostructure, we chose titania sol ($n_3$ = 1.78) and UV-curable resin ($n_4$ = 1.54) as dielectric materials, with Si served as the substrate. Based on the second suggestion in section 2.2, the number of dielectric layers was set to be 12. The OPD described in equation (3) was restricted to be 3150nm. Regarding layer thicknesses, the quarter-wave condition was adopted for each constituent 1D PC. On this occasion, three groups of material thicknesses were considered, denoted by "T-A" with {$400/4n_3$, $400/4n_4$, $650/4n_3$, $650/4n_4$}, "T-B" with {$450/4n_3$, $450/4n_4$, $600/4n_3$, $600/4n_4$}, and "T-C" with {$500/4n_3$, $500/4n_4$, $550/4n_3$, $550/4n_4$}, respectively. Fig. 4(b) shows the reflection spectra of the three heterostructures. A similar result can be obtained that the relative intensities of the three reflective bands can be tuned by taking different material thickness combinations.

A brief comparison between the two types of 1D PC construction can be made here. For conventional 1D PC architecture, there are some obvious advantages: high reflectance due to high refractive index contrast of dielectric materials, insensitivity of high reflective bands to change of substrate, and tunability of the distribution of high reflective bands. However, its disadvantages are also noticeable: thick layer thicknesses, inability to adjust relative intensities of reflectance between high reflectance bands, and narrow bandwidth. For the 1D heterostructure, the situation is totally different. Its striking merits include thinner layer thicknesses, adjustability of the relative intensities of reflectance between reflectance bands, and wider bandwidth. There are also notable demerits containing lower reflectance arising from low refractive index contrast of alternating layers, incapability of tuning the distribution of high reflective bands, and a strict requirement on the substrate (high refractive index). To conclude, both two kinds of 1D PC construction have their own distinct strengths and deficiencies, and appropriate configurations can be chosen according to practical needs.

3.3 Multiple high reflective bands towards structural colors

A universally acknowledged expression of describing 1D PCs with structural colors can be accomplished by converting their reflection spectra into CIE 1931 chromaticity coordinates. In the following context, a series of related calculations are conducted based on the before-mentioned four examples. We start from the conventional 1D PC architecture with yellow and blue double-mode reflective bands in the visible region. First, 21 values of $\lambda_{1st}$ were chosen ranging from 2100nm to 2500nm, with a uniform interval of 20nm. Positions of the two high reflective bands changed as the value of $\lambda_{1st}$ increased, leading to a shift of CIE coordinate. For each value of $\lambda_{1st}$, the theoretical reflection spectrum of corresponding conventional 1D PC on glass substrate was obtained. Then, 21 pairs of CIE coordinates were calculated from these reflection spectra, and the resulting CIE 1931 chromaticity diagram is illustrated in Fig. 5(a). Several pairs of the CIE coordinates are listed below: (0.2784,0.5216) (labeled as #1, $\lambda_{1st}$ = 2100nm), (0.3203,0.3964) (labeled as #6, $\lambda_{1st}$ = 2200nm), (0.3716,0.3108) (labeled as #11, $\lambda_{1st}$ = 2300nm), (0.4278,0.2850) (labeled as #16, $\lambda_{1st}$ = 2400nm), and (0.4235,0.3253) (labeled as #21, $\lambda_{1st}$ = 2500nm). Fig. 5(a) reveals that the CIE coordinate experienced a redshift, passing by the pure white light region.

Next, we move forward to 1D PC heterostructure with yellow and blue double-mode reflective bands in the visible region. To collect richer data for CIE coordinates, a wider range of material thickness combinations is designed with the OPD fixed at 2300nm. Specifically, 16 values of $\lambda_1$ were selected in a range from 400nm to 550nm, with a uniform interval of 10nm. Correspondingly, there would be 16 values of $\lambda_2$ decreasing uniformly from 750nm to 600nm. The reflectance of the two high reflective bands changed as the values of $\lambda_1$ and $\lambda_2$ varied, leading to a shift of CIE coordinate. For each pair of $\lambda_1$ and $\lambda_2$, the theoretical reflection spectrum of the corresponding 1D PC heterostructure on Si substrate was obtained. Then, 16 pairs of CIE coordinates were calculated from these reflection spectra, and the resulting CIE 1931 chromaticity diagram is demonstrated in Fig. 5(b). Several pairs of the CIE coordinates are listed below: (0.3303,0.2745) (labeled as #1, $\lambda_1$ = 400nm), (0.3630,0.3107) (labeled as #6, $\lambda_1$ = 450nm), (0.3914,0.3667) (labeled as #11, $\lambda_1$ = 500nm), and (0.4120,0.4128) (labeled as #16, $\lambda_1$ = 550nm). Fig. 5(b) shows that the CIE coordinate underwent a redshift, moving from the cold white light region to the warm white light region.

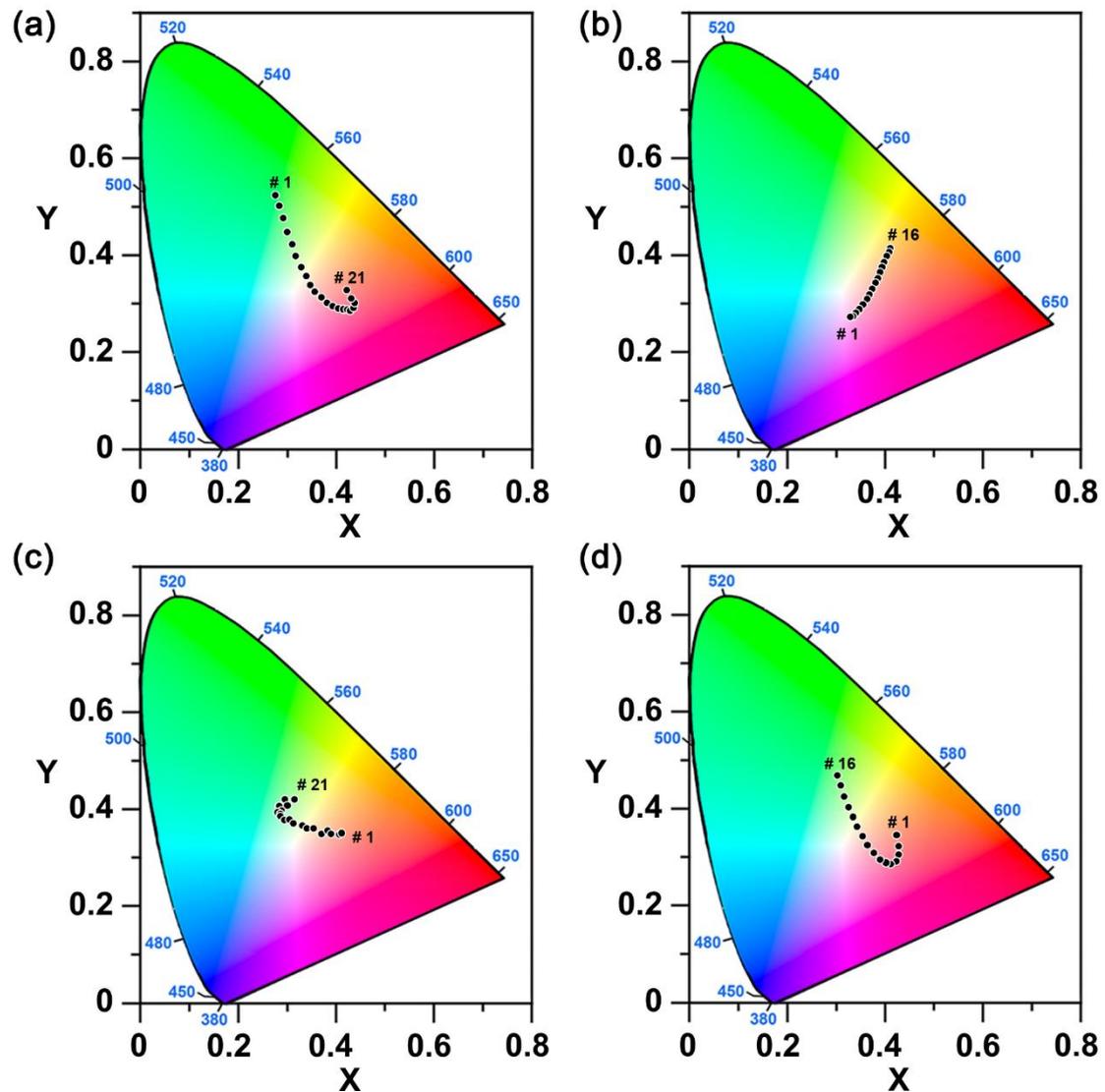

FIG. 5. CIE 1931 chromaticity coordinates of (a) conventional 1D PCs with double-mode reflective bands in the visible region. (b) 1D PC heterostructures with double-mode reflective bands in the visible region. (c) conventional 1D PCs with triple-mode

reflective bands in the visible region. (d) 1D PC heterostructures with triple-mode reflective bands in the visible region.

Currently, we proceed to consider conventional 1D PC architecture with red, green, and blue triple-mode reflective bands in the visible region. A similar calculation process can be directly applied here. First, 21 values of $\lambda_{1st}$ were chosen ranging from 2950nm to 3350nm, with a uniform interval of 20nm. For each value of $\lambda_{1st}$, the theoretical reflection spectrum of corresponding conventional 1D PC on glass substrate was obtained. Then, 21 pairs of CIE coordinates were calculated from these reflection spectra, and the resultant CIE 1931 chromaticity diagram is shown in Fig. 5(c). Several pairs of the CIE coordinates are listed below: (0.4111,0.3505) (labeled as #1, $\lambda_{1st}$ = 2950nm), (0.3840,0.3550) (labeled as #6, $\lambda_{1st}$ = 3050nm), (0.3139,0.3696) (labeled as #11, $\lambda_{1st}$ = 3150nm), (0.2833,0.3937) (labeled as #16, $\lambda_{1st}$ = 3250nm), and (0.3160,0.4179) (labeled as #21, $\lambda_{1st}$ = 3350nm). Fig. 5(c) illustrates that the CIE coordinate experienced a blue shift, with the trajectory mainly covered in the warm white light region.

Finally, we continue the study on 1D PC heterostructure with red, green, and blue triple-mode reflective bands in the visible region. A similar calculation process can be directly applied here. A broader collection of material thickness combinations is devised with the OPD fixed at 3150nm. Specifically, 16 values of $\lambda_1$ were picked in a range from 350nm to 500nm, with a uniform interval of 10nm. Correspondingly, there would be 16 values of $\lambda_2$ decreasing uniformly from 700nm to 550nm. For each pair of $\lambda_1$ and $\lambda_2$, the theoretical reflection spectrum of the corresponding 1D PC heterostructure on Si substrate was obtained. Then, 16 pairs of CIE coordinates were calculated from these reflection spectra, and the resultant CIE 1931 chromaticity diagram is depicted in Fig. 5(d). Several pairs of the CIE coordinates are listed below: (0.4241,0.3440) (labeled as #1, $\lambda_1$ = 350nm), (0.4024,0.2862) (labeled as #6, $\lambda_1$ = 400nm), (0.3441,0.3625) (labeled as #11, $\lambda_1$ = 450nm), and (0.3032,0.4686) (labeled as #16, $\lambda_1$ = 500nm). Fig. 5(d) indicates that the CIE coordinate underwent a blue shift, passing by the pure white light region.

Different from traditional CIE coordinate shifts caused by one major reflective band movement [28–31], the CIE coordinate shifts shown in Fig. 5 all exhibit certain tunability in the white light region, which can be mainly attributed to color mixing resulting from multiple reflective bands. The results in Figs. 5(a) and 5(c) indicate that triple-mode reflective bands maintain greater white-light stability than double-mode reflective bands when facing band position change. Figs. 5(b) and 5(d) reveal that band reflectance variation causes more complicated CIE coordinate performance for triple-mode reflective bands, and it is reasonable because three colors form an area in the CIE chromaticity diagram while two colors can only make a line. To conclude, both band position change and band reflectance variation contribute to shifting of CIE coordinate, with various effects on double-mode and triple-mode reflective bands.

## 4. Conclusion

In summary, we present two different methods to manipulate multiple reflective bands. The given examples indicate that positions of multiple reflective bands can be fine-tuned under certain theoretical guidance for both types of 1D PC construction. Besides, the relative intensities of reflectance between multiple reflective bands can be modulated by using the 1D PC heterostructure. Structural color is signified by converting reflection spectra into CIE coordinates, and the obtained results reveal that CIE coordinate shift occurs due to either band position change or band reflectance variation. The CIE coordinate shift behavior is also influenced by the number of multiple reflective bands. The two approaches reported in this work can also be extended to the manipulation of multi-mode reflective bands in the infrared and ultraviolet regions.

## ACKNOWLEDGEMENTS

This work was supported by National Natural Science Foundation of China (grant number 21973033) and the Fundamental Research Funds for the Central Universities (grant number 2018ZD13).